# Data-Driven Topology Optimization with Multiclass Microstructures using Latent Variable Gaussian Process

Liwei Wang[a,b]     Siyu Tao[b]     Ping Zhu[a]     Wei Chen[b]

a. The State Key Laboratory of Mechanical System and Vibration,
Shanghai Key Laboratory of Digital Manufacture for Thin-Walled Structures,
School of Mechanical Engineering,
Shanghai Jiao Tong University
Shanghai, P.R. China

b. Dept. of Mechanical Engineering
Northwestern University
Evanston, IL, USA

**ABSTRACT**

The data-driven approach is emerging as a promising method for the topological design of multiscale structures with greater efficiency. However, existing data-driven methods mostly focus on a single class of microstructures without considering multiple classes to accommodate spatially varying desired properties. The key challenge is the lack of an inherent ordering or "distance" measure between different classes of microstructures in meeting a range of properties. To overcome this hurdle, we extend the newly developed latent-variable Gaussian process (LVGP) models to create multi-response LVGP (MR-LVGP) models for the microstructure libraries of metamaterials, taking both qualitative microstructure concepts and quantitative microstructure design variables as mixed-variable inputs. The MR-LVGP model embeds the mixed variables into a continuous design space based on their collective effects on the responses, providing substantial insights into the interplay between different geometrical classes and material parameters of microstructures. With this model, we can easily obtain a continuous and differentiable transition between different microstructure concepts that can render gradient information for multiscale topology optimization. We demonstrate its benefits through multiscale topology optimization with aperiodic microstructures. Design examples reveal that considering multiclass microstructures can lead to improved performance due to the consistent load-transfer paths for micro- and macro-structures.

**Keywords**: multiscale topology optimization, Gaussian process, mixed variables, multiclass, data-driven design

## 1. INTRODUCTION

The rapid development of additive manufacturing has made it possible to fabricate components with rather complex structures, enabling greater freedom for structure design. Along with the enhancement in manufacturing capability, there is a growing interest in topology optimization (TO) for multi-scale structure design [1]. Specifically, the layout of materials for the structure is optimized in the macro-scale while the local material properties are controlled by varying the configuration and/or material constituents of microstructures in the micro-scale. By combining micro- and macro-structure designs, the full structure is expected to achieve better functionalities than the single-scale design [2]. However, multiscale structure design faces enormous computational challenges due to the infinite dimensionality of geometrical designs and the nested micro- and macro-scale analyses. To address this computational challenge, this research aims to develop a data-driven approach that can significantly expedite multiscale TO through a novel mixed-variable Gaussian process (GP) modeling technique, which allows the concurrent exploration of microstructure concepts and the associated geometric and/or material variables.

Following the pioneering work of Rodrigues et al. [3], various TO methods have been developed for the design of multi-scale structures. A relatively direct type of methods is to assume a periodically assembled full structure and then perform the optimization in two scales separately [4-6] or concurrently [7, 8]. While these methods are efficient, using a single type of microstructure significantly reduces the computational requirement at the cost of the suboptimal solution and is not able to accommodate spatially varying property requirements. In contrast, Xia et al. [9, 10] proposed an $FE^2$-based method to enable element-wise microstructure design. Although their method can provide greater design freedom, the resultant computation cost is excessive. As a compromise between computation cost and design freedom, several concurrent design methods reduce the design space by dividing a full structure into a small number of subregions with the same microstructure [11-16]. As a result, the original fully aperiodic design is replaced by clusters of periodic designs, which can greatly accelerate the optimization process. Nevertheless, the full structure design is confined to a fixed number of microstructures. There is a need for a multi-scale TO algorithm with a high efficiency while offering a large design freedom for aperiodic microstructures.

In recent literature, the data-driven approach has shown promises to address the aforementioned challenges for multiscale TO. Early developments of the data-driven approach focused on a single class of microstructures (topology) and obtain microstructures with different solid material volume fractions by changing the predefined geometric parameters (e.g., rod thickness). The iterative evaluation of effective properties in the nested microstructure design is replaced by a regression function of the relation between volume fraction and the precomputed properties, such as exponential function [17], polynomial [18-20], kriging model [21], neural network [22] and diffuse approximation [23]. Full structures with partially varying porosity can be obtained efficiently by this single-class framework. However, these methods lead to suboptimal solutions since only a single predefined class of microstructure is used in the whole design process.

To enable the consideration of multiple classes of microstructures in the data-driven design, Wang et al. [24, 25] proposed a sophisticated parameterization method for selected classes of truss microstructures by controlling the aspect ratio. However, this parameterization technique is difficult to be generalized to other microstructures with different topologies. Alternatively, the complex shapes of microstructures can be represented by some reduced-order shape descriptors, such as the latent variables for the deep generative model [26] or the Laplace-Beltrami spectrum [27, 28], to enable machine learning for accelerating the design process. Nevertheless, these methods extract descriptors only based on geometries but not properties. As a result, the descriptors of microstructures may be hard to interpret and unnecessarily high-dimensional.

Overall, physics-based multiscale TO methods are generally time-consuming while existing data-driven approaches have difficulties in handling multiple classes of microstructures. There is a need for an efficient data-driven multiscale design method that can incorporate multiple classes of microstructures to provide spatially variant microstructure designs for improved structural performance.

## 2. OVERVIEW OF THE PROPOSED FRAMEWORK

We view the multiscale TO as a concurrent macro- and micro-structure design as shown in Figure. 1. In this concurrent design process, design variables for the microstructure fall into two categories: *quantitative* (e.g., porosity, element material property) and *qualitative* variables (e.g., class of microstructure and material composition). These two types of variables are coupled together to determine the homogenized stiffness matrix of a microstructure represented by a number of independent matrix components, e.g., $C_{11}, C_{12}, C_{22}, C_{66}$ for 2D orthotropic microstructures. Therefore, the structure-property relation of a microstructure can be considered as a multi-response physical model with mixed-type variables as inputs. Note that the qualitative variable normally does not have a distance measurement between different levels, which is different from integer or real-value variables with intrinsic distance metric. As a result, no neighboring information or gradient value can be obtained from the qualitative variable, imposing a major challenge for surrogate modeling and optimization.

With this in mind, we propose to construct a unified and continuous design space that allows the concurrent exploration of microstructure concepts and the associated geometric and/or material parameters. Specifically, a new surrogate modeling method, multi-response latent variable Gaussian process (MR-LVGP) modeling, is proposed by generalizing our recently proposed LVGP model to enable the Gaussian process modeling for datasets with multiple responses and mixed-variable inputs. This MR-LVGP model is created using the multiclass microstructure libraries with precomputed properties, surrogating the structure-property relations of microstructures. The special feature of the fitted MR-LVGP model is that the unordered classes of microstructures can be mapped into a continuous and well-organized latent space based on their effects on responses. In this study, a latent variable is defined to be an underlying variable that is essential for the physical model description but not directly observed from the input data, which can only be obtained by statistical inference [29]. By varying the continuous latent variables, the stiffness matrix (properties) predicted by MR-LVGP models will have a smooth transition between different classes of microstructures. As a result, the neighboring information and the gradient of properties with respect to latent variables can be obtained for the multiscale topology optimization.

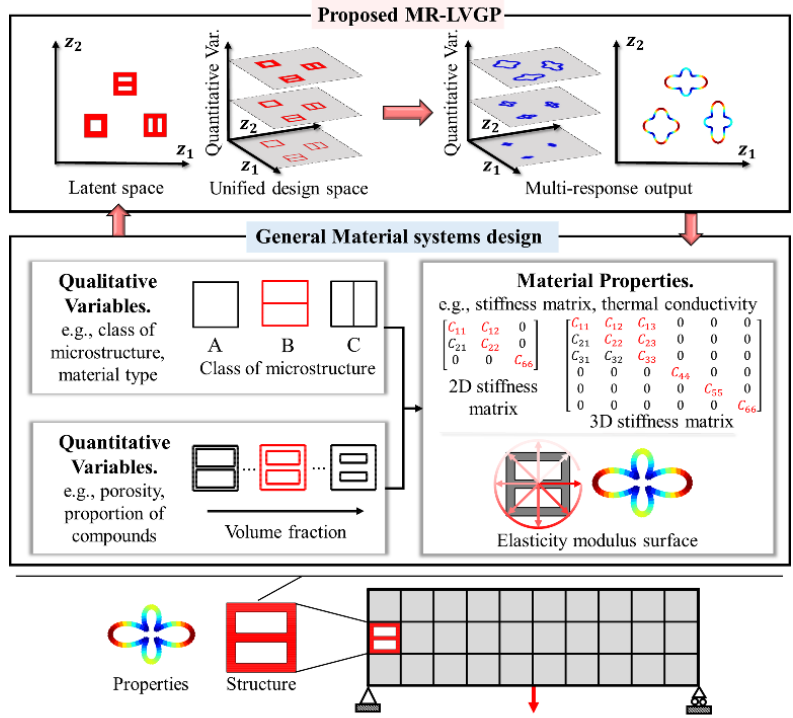

**FIGURE 1**: The proposed data-driven design framework for assembling microstructure designs described by both qualitative and quantitative variables. The independent components in the stiffness matrix are considered as multiple responses. The stiffness matrix is visualized by rotating the stiffness matrix to obtain the modulus surface, with the distance from the center representing the magnitude of $C_{11}$ value. The color code on the modulus surface is used to better visualize the distribution of $C_{11}$ value in different directions.

The latent space underlying qualitative variables is then combined with the quantitative variables, e.g., volume fraction, to form a unified design space for the multiscale TO of the full structure. As a result, the variations of microstructure classes and their associated quantitative parameters simply represent different moving directions in the unified space. With MR-LVGP models that connect micro- and macro-scale analyses, matured density-based TO methods, e.g. SIMP method, can be directly applied to the latent variables. In fact, the MR-LVGP model can be seen as a generalized interpolation scheme for the density-based TO, taking the class of microstructure into consideration. Specifically, we add the latent variables as extra design variables for the SIMP method and use MR-LVGP models to provide stiffness matrix as well as its associated sensitivity for the optimization in each iteration. Following the technique used in SIMP to avoid intermediate densities, we propose to add a penalization term to the stiffness matrix based on the inherent distance in the latent space, driving the design process to converge to predefined classes of microstructures. A closely related work is the topology optimization design based on Shared-GP modeling proposed by Xing et al. [30], constructing a shared latent space to represent correlations within and across multiple design spaces. However, their work only focuses on the single-scale macrostructure design and does not consider any categorical design variable. The remaining paper begins with a brief review of our LVGP modeling method (Section 3) and its extension to MR-LVGP modeling based on multi-response datasets (Section 4). A library that consists of multiclass microstructures is constructed for both 2D and 3D cases. MR-LVGP models are fitted to each library to obtain a continuous latent space for different classes of microstructures (Section 5). The fitted MR-LVGP models are incorporated into the TO algorithm with a penalization technique, enabling the multiscale design with multiple types of microstructures. The advantages of considering multiple microstructure types are demonstrated through both 2D and 3D multiscale design cases (Section 6). Finally, our conclusions are drawn in Section 7.

## 3. REVIEW OF LVGP MODELING

GP modeling for functions with continuous input variables has been well established in past decades [31, 32]. The core of GP modeling is to regard responses at different inputs as realizations of jointly distributed Gaussian random variables, and the correlations between the Gaussian random variables depend on the distances between the inputs. However, it is not straightforward to extend this method to functions with mixed-variable inputs, as the distances between the levels of the qualitative variables are rarely well-defined.

A number of works to address this challenge have been reported [33-38], yet most methods rely on simplifications in the covariance structure based on specialized domain knowledge or heuristic assumptions [39]. In contrast, we recently developed a novel LVGP modeling method to enable GP modeling with any standard GP correlation function for mixed-variable datasets in a straightforward and computationally stable manner [40]. This method has been successfully applied to the design of different material systems, such as the light-absorbing quasi-random solar cell [41] and insulating nanocomposites [42]. We have shown its greater flexibility and superior predictive performance over existing alternatives across a wide variety of problems. Moreover, LVGP can provide a continuous and meaningful embedding for qualitative variables, which is highly desirable for gradient-based optimization in this study.

As illustrated in Figure. 2, the key intuition of LVGP modeling is that the effects of any qualitative factor on a quantitative response must always be due to some underlying quantitative physical input variables $\boldsymbol{V} = \{v_1, v_2, \ldots, v_n\}, \in \mathrm{R}^n$. Although these underlying variables can be extremely high-dimensional, their collective effects can be represented approximately by a function $g(v_1, v_2, \ldots, v_n)$ residing in a low-dimensional manifold, whose local manifold coordinates can be used as reduced-dimensional descriptors for different qualitative variable combinations [43, 44]. In our case, the underlying physical variables for different classes of microstructures can be different types of shape descriptors, e.g., the pixel/voxel matrix, nodes of boundary splines, and some other geometrical parameters. These underlying variables can be considered as a set on a low-dimensional manifold in the sense that a feasible structure will impose some implicit constraints to these parameters (e.g., the solid domain should be connected). Based on this insight, LVGP modeling assumes a latent space (e.g., two dimensional $z_1$-$z_2$ space in Figure. 2) that corresponds to local coordinates on the manifold and maps the levels of the qualitative variables (e.g. level 1, 2, 3 in Figure. 2) to some locations in the latent space (e.g. the discrete points in the $z_1$-$z_2$ space in Figure. 2). The existing distance-based correlations can then be applied to these levels through their latent variables in GP modeling. It should be noted that the mapping is constructed directly between the qualitative variables and the latent variables without the need to identify any underlying quantitative physical input for qualitative variables. The corresponding latent variables are treated as undetermined parameters and estimated efficiently by maximizing the likelihood function of the LVGP model, which will be illustrated in the remaining part of this section.

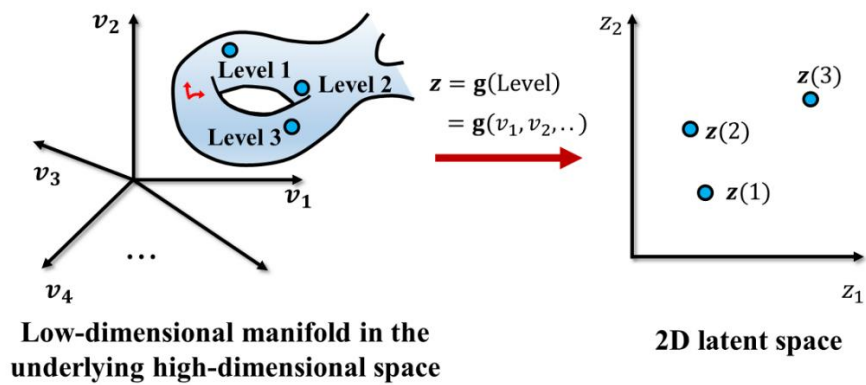

**FIGURE 2**: An illustration of the mapping from the high-dimensional underlying quantitative variables to the 2D latent variables for a qualitative variable with three discrete levels, the red axes on the manifold represent a local coordinate system.

Consider a single-response computer simulation model $y(\boldsymbol{w})$ with input $\boldsymbol{w} = [\boldsymbol{x}^T, \boldsymbol{t}^T]^T$ containing both quantitative variables $\boldsymbol{x} = [x_1, x_2, \dots, x_p]^T \in R^p$ and qualitative variables $\boldsymbol{t} = [t_1, t_2, \dots, t_q]^T$, with the $j^{th}$ qualitative factor $t_j \in \{1, 2 \dots, l_j\}$. Herein, $l_j \in N^+$ is the total number of levels for the $j^{th}$ qualitative factor $t_j$. Assume that each $t_j$ is mapped to a $g$-dimensional latent vector $\boldsymbol{z}_j(t_j) = [z_{j,1}(t_j), \dots, z_{j,g}(t_j)]^T \in R^g$. Denote the transformed input vector as $\boldsymbol{s} = [\boldsymbol{x}^T, \boldsymbol{z}(\boldsymbol{t})^T]^T \in R^{p+q \times g}$, where $\boldsymbol{z}(\boldsymbol{t}) = [\boldsymbol{z}_1(t_1)^T, \dots, \boldsymbol{z}_q(t_q)^T]^T$. The standard GP model can then be modified as (using a constant mean function)

$$Y(\boldsymbol{s}) = \beta + G(\boldsymbol{s}), \tag{1}$$

where $\beta$ is the constant mean and $G(\cdot)$ is a zero-mean Gaussian process with its correlation defined as

$$cov\big(G(\boldsymbol{s}), G(\boldsymbol{s}')\big) = c(\boldsymbol{s}, \boldsymbol{s}') = \sigma^2 r(\boldsymbol{s}, \boldsymbol{s}'), \tag{2}$$

where $\sigma^2$ is the prior variance, $c(\cdot,\cdot)$ is the covariance function, and $r(\cdot,\cdot)$ is the correlation function. Among numerous existing correlation functions, the Gaussian correlation function is commonly used:

$$r(s, s') = \exp\left\{-(x-x')^T \Phi(x-x') - (z(t)-z(t'))^T(z(t)-z(t'))\right\}, \quad (3)$$

where $\Phi = diag(\phi)$ and $\phi = [\phi_1, \phi_2, ..., \phi_p]^T$ are scaling parameters to be estimated. Because the latent variables $z(t)$ are to be estimated as hyperparameters, their scaling parameters are set to ones to avoid over-parameterization.

For a size-$n$ training dataset with inputs $W = [w^{(1)}, w^{(2)}, ..., w^{(n)}]^T$ and outputs $y = [y^{(1)}, y^{(2)}, ..., y^{(n)}]^T$, the corresponding log-likelihood function is

$$L_{ln}(Z, \phi, \beta, \sigma^2) = -\frac{n}{2}\ln(\sigma^2) - \frac{1}{2}\ln|R(Z,\phi)| - \frac{1}{2\sigma^2}(y-1\beta)^T R(Z,\phi)^{-1}(y-1\beta), \quad (4)$$

where $\ln(\cdot)$ is the natural logarithm, $\mathbf{1}$ is an $n \times 1$ vector of ones, $R$ is the $n \times n$ correlation matrix with $R_{ij} = r(s^{(i)}, s^{(j)})$ for $i, j = 1, ..., n$, and $Z = \cup_{i=1}^{q}\{z_i(1), ..., z_i(l_i)\}$ is the set of mapped latent variable values for all the levels of the qualitative variables.

By setting derivatives of (4) with respect to $\beta$ and $\sigma^2$ to be zero, we obtain maximum likelihood estimates (MLEs) for $\beta$ and $\sigma^2$:

$$\hat{\beta} = \frac{\mathbf{1}^T R^{-1} y}{\mathbf{1}^T R^{-1} \mathbf{1}}, \quad (5)$$

$$\hat{\sigma}^2 = \frac{1}{n}(y-1\hat{\beta})^T R^{-1}(y-1\hat{\beta}). \quad (6)$$

After substituting (5) and (6) into (4), the estimates of $\hat{Z}$ and $\hat{\phi}$ can be obtained by minimizing the negative log-likelihood function (ignore constant terms):

$$[\hat{Z}, \hat{\phi}] = \underset{Z, \phi}{\text{argmin}}\ n\ln(\hat{\sigma}^2) + \ln(|R|). \quad (7)$$

This minimization problem can be solved with various mature optimization algorithms. The prediction for the response $y(w^*)$ can then be made at $w^*$ by

$$\hat{y}(w^*) = \hat{\beta} + r^T R^{-1}(y - 1\hat{\beta}), \quad (8)$$

where $r = [r(s^*, s^{(1)}), r(s^*, s^{(2)}), ..., r(s^*, s^{(n)})]^T$.

In this way, the qualitative variables can be transformed into continuous latent variables according to their effects on output. For more detailed illustrations and implementation of the LVGP modeling, readers are referred to [40].

## 4. MULTI-RESPONSE LVGP MODELING

The original LVGP modeling was proposed only for single-response computer simulation models. However, the responses of material properties are often multi-dimensional. For example, in our case, the responses are the independent components in the stiffness matrix, e.g., $C_{11}$ and $C_{12}$, for the mechanical constitutive relations, which would be four-dimensional for 2D orthotropic microstructures and nine-dimensional for 3D microstructures. A naive method is to fit a single-response (SR) LVGP model for each output separately and transform the qualitative variables into different latent spaces for each output. However, this is not the most efficient way to handle qualitative variables and it is more desirable to obtain unified latent variables for them by considering all the responses and their correlations. Herein, we follow the procedure in [45] to extend the LVGP to multi-response cases.

Consider a multi-response computer simulation model $\boldsymbol{y}(\boldsymbol{w})$ with output $\boldsymbol{y} = [y_1, y_2, \dots, y_{N_{op}}]^T \in R^{N_{op}}$ and input $\boldsymbol{w} = [\boldsymbol{x}^T, \boldsymbol{t}^T]^T$. Assume the prior model for the outputs is

$$\boldsymbol{Y}(\boldsymbol{s}) = \boldsymbol{B}^T \boldsymbol{h}(\boldsymbol{w}) + \boldsymbol{G}(\boldsymbol{s}), \tag{9}$$

where $\boldsymbol{s}$ has the same definition as in Section 3, $\boldsymbol{h}(\cdot)$ is the prior mean function composed of a vector of given regression functions $[h_1(\cdot), h_2(\cdot), \dots, h_v(\cdot)]^T$, $\boldsymbol{B}$ is $[\boldsymbol{\beta}_1, \boldsymbol{\beta}_2, \dots, \boldsymbol{\beta}_{N_{op}}]$, a matrix of unknown regression coefficients with $\boldsymbol{\beta}_i = [\beta_{1,i}, \beta_{2,i}, \dots, \beta_{v,i}]^T$, and $\boldsymbol{G}$ is $[G_1, G_2, \dots, G_{N_{op}}]^T$, a multi-response stationary Gaussian process with zero mean values and separable covariance structure. In this section, we will use $\boldsymbol{s}$ as the mapped input of $\boldsymbol{w}$ without further notice. The covariance matrix between the outputs at any given pair of inputs is

$$\boldsymbol{cov}(\boldsymbol{G}(\boldsymbol{s}), \boldsymbol{G}(\boldsymbol{s}')) = \boldsymbol{\Sigma} \cdot r(\boldsymbol{s}, \boldsymbol{s}'), \tag{10}$$

or written component-wise as

$$cov\left(G_i(\boldsymbol{s}), G_j(\boldsymbol{s}')\right) = \Sigma_{ij} \cdot r(\boldsymbol{s}, \boldsymbol{s}'), \tag{11}$$

where $\Sigma_{ij}$ is the corresponding entry of an unknown nonspatial $N_{op} \times N_{op}$ covariance matrix $\boldsymbol{\Sigma}$, and $r(\cdot, \cdot)$ is a spatial correlation function with the same definition as in (3). Compared with the covariance definition in LVGP modeling, the covariance for the MR-LVGP model becomes a matrix with each entry composed of the spatial correlation between different input vectors and an extra term $\Sigma_{ij}$ capturing the covariance between the pair of response variables.

Consider a training dataset for MR-LVGP with input data $\mathbf{W} = [\mathbf{w}^{(1)}, \mathbf{w}^{(2)}, \ldots, \mathbf{w}^{(n)}]^T$ and observed response data $\mathbf{D} = [\mathbf{y}^{(1)}, \mathbf{y}^{(2)}, \ldots, \mathbf{y}^{(n)}]^T$, the corresponding log-likelihood is (with constants dropped)

$$L_{ln}(\mathbf{Z}, \boldsymbol{\phi}, \mathbf{B}, \boldsymbol{\Sigma}) = -\frac{n}{2}\ln|\boldsymbol{\Sigma}| - \frac{q}{2}\ln|\mathbf{R}| - \frac{1}{2}vec(\mathbf{D} - \mathbf{H}\mathbf{B})^T(\boldsymbol{\Sigma}\otimes\mathbf{R})^{-1}vec(\mathbf{D} - \mathbf{H}\mathbf{B}), \quad (12)$$

where $vec(\cdot)$ converts a matrix to a column vector by stacking the columns of a matrix, $\otimes$ denotes the Kronecker product, $\mathbf{H}$ is $[\mathbf{h}(\mathbf{w}^{(1)}), \ldots, \mathbf{h}(\mathbf{w}^{(n)})]^T$, a matrix containing all the basis function values for input data, and $\mathbf{R}$ is an $n \times n$ correlation matrix with $R_{ij} = r(\mathbf{s}^{(i)}, \mathbf{s}^{(j)})$ for $i, j = 1, \ldots, n$.

Noting that $(\boldsymbol{\Sigma}\otimes\mathbf{R})^{-1} = \boldsymbol{\Sigma}^{-1}\otimes\mathbf{R}^{-1}$, we can obtain the MLEs for parameters $\mathbf{B}$ and $\boldsymbol{\Sigma}$ by following a similar practice to that in Section 3 [46]:

$$\widehat{\mathbf{B}} = (\mathbf{H}^T\mathbf{R}^{-1}\mathbf{H})^{-1}\mathbf{H}^T\mathbf{R}^{-1}\mathbf{D}, \quad (13)$$

$$\widehat{\boldsymbol{\Sigma}} = \frac{1}{n}(\mathbf{D} - \mathbf{H}\widehat{\mathbf{B}})^T\mathbf{R}^{-1}(\mathbf{D} - \mathbf{H}\widehat{\mathbf{B}}). \quad (14)$$

The MLEs for $\mathbf{Z}$ and $\boldsymbol{\phi}$ can be obtained by maximizing the log-likelihood function in (12) after substituting $\mathbf{B}$ and $\boldsymbol{\Sigma}$ with $\widehat{\mathbf{B}}$ and $\widehat{\boldsymbol{\Sigma}}$. The prediction for the response $\mathbf{y}(\mathbf{w}^*)$ can then be made at $\mathbf{w}^*$ by

$$\widehat{\mathbf{y}}(\mathbf{w}^*) = \widehat{\mathbf{B}}^T\mathbf{h}(\mathbf{w}^*) + \mathbf{r}^T\mathbf{R}^{-1}(\mathbf{D} - \mathbf{H}\widehat{\mathbf{B}}), \quad (15)$$

where $\mathbf{r} = [r(\mathbf{s}^*, \mathbf{s}^{(1)}), r(\mathbf{s}^*, \mathbf{s}^{(2)}), \ldots, r(\mathbf{s}^*, \mathbf{s}^{(n)})]^T$.

## 5. MR-LVGP MODELING FOR MULTICLASS MICROSTRUCTURES

Our study aims to demonstrate the benefits of MR-LVGP modeling in the data-driven multiscale TO with aperiodic microstructures. In this section, we first construct libraries containing multiple microstructure patterns (classes) for both 2D and 3D cases. The MR-LVGP modeling method is then applied to these libraries, mapping different types of microstructures into a latent space for the multiscale TO in Section 6.

### 5.1. Construction of Microstructure Libraries

To accommodate different stress distributions under different loading conditions in a multiscale structure, the desired microstructure libraries should contain diverse structures to meet a range of mechanical properties. In the 2D library, for the purpose of illustration, we focus on orthotropic microstructures and use the six classes shown in Figure. 3. In each class, the structure consists of a set of rods with the same thickness. Therefore, when the volume fraction of the microstructure is given, the structure is fully determined. All the microstructures are orthotropic, and classes A through D also have cubic symmetry, which means that their mechanical properties are the same in $x$ and $y$ directions. In contrast, classes E and F are stiffer in the $x$ and $y$ directions than the other direction, respectively. The difference between each class' effective stiffness matrix can also be illustrated through their homogenized elasticity modulus surfaces shown in the figure. The overall 2D library will require four independent components (responses) to fully represent the stiffness matrix, which are marked in red in Figure. 1 and represented by the unique markers of the schematic representative stiffness matrix in Figure. 3.

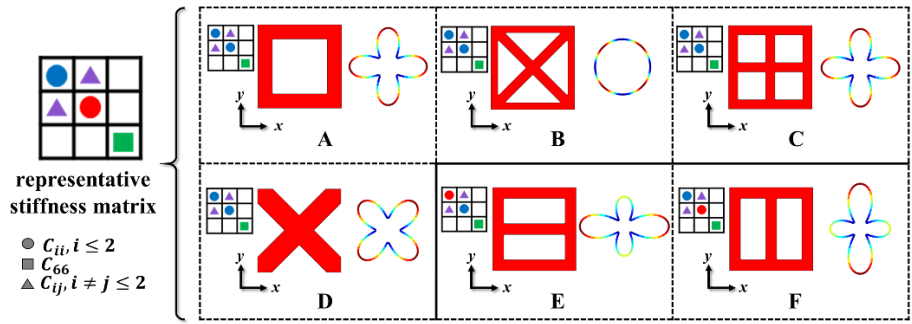

**FIGURE 3**: Different classes of microstructures for the 2D library. In each block, the representative structure is shown on the left while its homogenized elasticity modulus surface is on the right. The small $3 \times 3$ matrix is a schematic representation of the 2D effective stiffness matrix shown in Figure. 1. Different types of modulus are represented by different shapes, and independent components are represented by markers with unique combinations of shape and color.

For the 3D library, all 14 classes of microstructures are orthotropic as shown in Figure. 4, among which classes A through H have extra cubic symmetry. Therefore, the whole 3D library generally requires nine independent components (responses) to describe the 3D stiffness matrix, which are also marked in red in Figure. 1 and represented by the unique markers of the schematic representative stiffness matrix in Figure. 4. Similar to those in the 2D library, the 3D microstructures can be parameterized by the thickness of the thinnest rod, which can be determined for a given volume fraction. Note that another benefit of using these microstructures is that they are designed to be connected with each other. This feature can avoid the possible boundary compatibility issue in the macro-scale design.

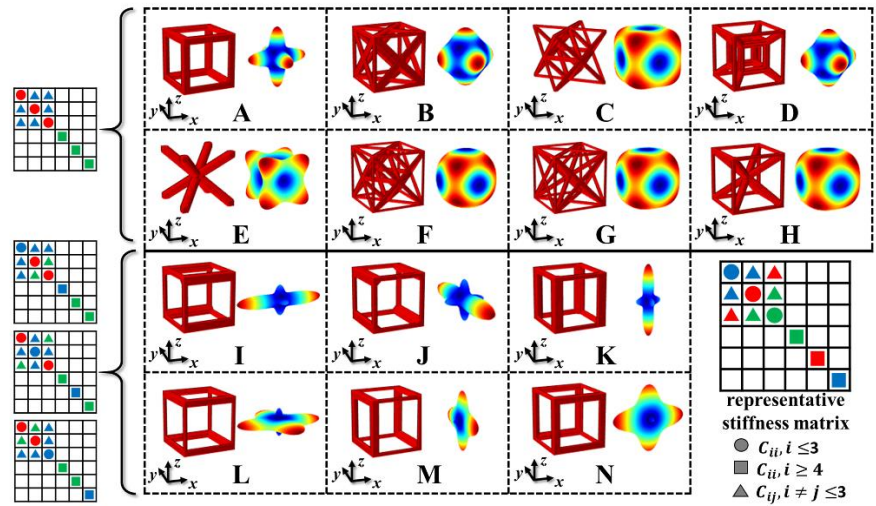

**FIGURE 4**: Different classes of microstructures for the 3D library. Microstructures in the first two rows have cubic symmetry. Classes I through K have thicker rods in $x$, $y$ and $z$ direction, respectively. Classes L through N have thicker rods in $x$-$y$, $y$-$z$ and $x$-$z$ directions, respectively. In each block, the representative structure is shown on the left while its homogenized elastic modulus surface is on the right. The small $6 \times 6$ matrix is a schematic representation of the 3D effective stiffness matrix corresponding to the 3D effective stiffness matrix shown in Figure. 1. Different types of modulus are represented by different shapes, and independent components are represented by markers with unique combinations of shape and color.

In this paper, the 2D microstructure is represented by a $100 \times 100$ pixel matrix while the 3D microstructure is discretized by a $50 \times 50 \times 50$ voxel cube. To construct multiclass libraries, we sample different microstructures for each class by uniformly varying the volume fraction (adjusted by the rod thickness). Based on our empirical study, the total numbers of microstructures are 120 for the 2D library and 261 for the 3D library. The stiffness matrices of the microstructures have a monotonic increase with small nonlinearity when the volume fraction is increased. Therefore, a satisfying prediction ability can be obtained with a mid-size dataset. In this study, the effective stiffness matrices for these microstructures are calculated by numerical homogenization [47].

**5.2. Construction of Continuous Latent Space by MR-LVGP Modeling**

Within the constructed libraries, there are two variables that control the structure of a microstructure, i.e. volume fraction and the qualitative class of microstructure, as shown in Figure. 5.

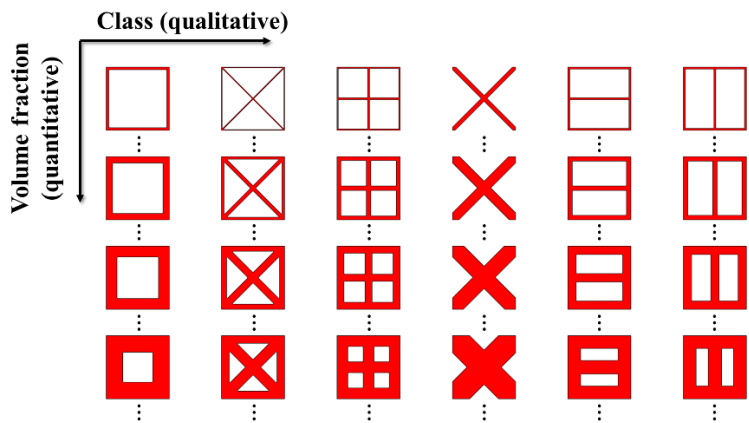

**FIGURE 5**: Two input variables for the library of microstructures. We use 2D microstructures for illustration.

We take the volume fraction $\rho$ as a quantitative input, class of microstructures $t$ as a qualitative input, and a vector of the independent components in the stiffness matrix $Y$ as a multi-dimensional response for the MR-LVGP modeling. Specifically, $Y$ is $[C_{11}, C_{12}, C_{22}, C_{66}]^T$ for 2D microstructures and $[C_{11}, C_{12}, C_{13}, C_{22}, C_{23}, C_{33}, C_{44}, C_{55}, C_{66}]^T$ for 3D microstructures. We adopt a 2D latent space for MR-LVGP modeling, which is reported in [40] to be sufficient for most physical problems. Constant mean functions are used for the MR-LVGP model, i.e. $h_i(w) = 1$ in (9).

As illustrated in the last subsection, we include 120 and 261 microstructures for 2D and 3D libraries, respectively, with uniformly sampled volume fraction for each class of microstructure. To validate the accuracy of MR-LVGP models for this problem, we randomly divide the dataset into training (80%) and test (20%) datasets for both 2D and 3D cases. An MR-LVGP model with 2D latent space is fitted to the training dataset and validated on the test dataset, which is repeated 10 times with random divisions of training/test datasets. To study the influence of the dimensionality of output and latent variables, we also train a single-response LVGP model for each property to obtain an assembled SR-LVGP model and MR-LVGP models with 3D and 4D latent space. The results are shown in Tables 1 and 2. While smaller values of the means and variances of MSEs mean more accurate predictions, these results show that MR-LVGP models with 2D latent space perform well on both 2D and 3D datasets, even though they involve complex behavior with high-dimensional outputs. MR-LVGP models with 3D or 4D latent spaces bring negligible improvements and even less accurate results in a few cases. We also perform the Mantel test [48] with $10^4$ permutations for distance matrices between latent vectors of microstructure classes under different dimensions. The result shows that the distance matrices in 2D latent space are highly correlated with those in 3D and 4D latent space ($r_M \geq 0.9214$, $p<0.01$ for 2D database; $r_M \geq 0.8915$, $p<10^{-6}$ for 3D database). Therefore, 2D latent space is enough to encode the information on the correlation between mechanical responses of different classes. This observation substantiates our previous finding in [40] that 2D latent space is sufficient for most physical models. In addition, MR-LVGP models retain good predictive capability with much lower dimensionalities of the latent spaces than assembled single-response (A-SR) LVGP models. For example, there are two latent variables associated with each of the nine properties for the 3D library, resulting in an $18D$ assembled latent space for the assembled model with highly correlated latent axes and sparsely distributed latent variables. High-dimensionality and sparsity of the ensemble latent space of single-response LVGP models will pose challenges to the data-driven multiscale design. Therefore, we will use the MR-LVGP model with 2D latent space in the remaining part.

**Table 1** MSE errors for MR-LVGP and assembled SR-LVGP models fitted for the 2D library. The mean and variance are calculated over 10 random repetitions.

|  | Mean of MSE | | | | Variance of MSE ($\times 10^{-6}$) | | | |
|---|---|---|---|---|---|---|---|---|
| Model | MR | MR | MR | A-SR | MR | MR | MR | A-SR |
| Dim. of $\mathbf{z}$ | 2 | 3 | 4 | 4×2 | 2 | 3 | 4 | 4×2 |
| $C_{11}$ | 0.0011 | 0.0009 | 0.0008 | 0.0004 | 2.0408 | 1.0937 | 1.1940 | 0.2051 |
| $C_{12}$ | 0.0002 | 0.0002 | 0.0002 | 0.0001 | 0.0627 | 0.0357 | 0.0381 | 0.0271 |
| $C_{22}$ | 0.0014 | 0.0013 | 0.0013 | 0.0011 | 1.8928 | 2.0425 | 1.7767 | 2.8194 |
| $C_{66}$ | 0.0002 | 0.0001 | 0.0002 | 0.0001 | 0.0481 | 0.0215 | 0.0308 | 0.0105 |

**Table 2** MSE errors for MR-LVGP and assembled SR-LVGP models fitted for the 3D library. The mean and variance are calculated over 10 random repetitions.

|  | Mean of MSE | | | | Variance of MSE ($\times 10^{-5}$) | | | |
|---|---|---|---|---|---|---|---|---|
| Model | MR | MR | MR | A-SR | MR | MR | MR | A-SR |
| Dim. of $\mathbf{z}$ | 2 | 3 | 4 | 9×2 | 2 | 3 | 4 | 9×2 |
| $C_{11}$ | 0.0190 | 0.0173 | 0.0182 | 0.0048 | 4.8665 | 1.4299 | 0.0287 | 0.0324 |
| $C_{12}$ | 0.0027 | 0.0028 | 0.0028 | 0.0006 | 0.0846 | 0.0640 | 0.0000 | 0.0307 |
| $C_{13}$ | 0.0026 | 0.0027 | 0.0026 | 0.0004 | 0.1037 | 0.1229 | 0.0123 | 0.0013 |
| $C_{22}$ | 0.0188 | 0.0227 | 0.0242 | 0.0040 | 1.8733 | 4.2234 | 0.3256 | 0.0019 |
| $C_{23}$ | 0.0026 | 0.0028 | 0.0026 | 0.0004 | 0.1290 | 0.0869 | 0.0053 | 0.0009 |
| $C_{33}$ | 0.0182 | 0.0256 | 0.0202 | 0.0041 | 5.6794 | 1.3148 | 1.2126 | 0.0519 |
| $C_{44}$ | 0.0017 | 0.0017 | 0.0019 | 0.0002 | 0.0407 | 0.0230 | 0.0063 | 0.0002 |
| $C_{55}$ | 0.0018 | 0.0019 | 0.0018 | 0.0002 | 0.0739 | 0.0125 | 0.0054 | 0.0000 |
| $C_{66}$ | 0.0019 | 0.0019 | 0.0021 | 0.0001 | 0.0521 | 0.0030 | 0.0000 | 0.0000 |

Figure 6 presents the 2D latent spaces for the MR-LVGP models obtained for 2D and 3D cases. The latent space is only constructed for the qualitative variable (class of microstructures). From the result, we can see that MR-LVGP models capture well the correlation between the mechanical responses of different classes by their distances in the 2D latent space. The larger the distance in the latent space between two classes of microstructures, the weaker the correlation between them in terms of their responses.

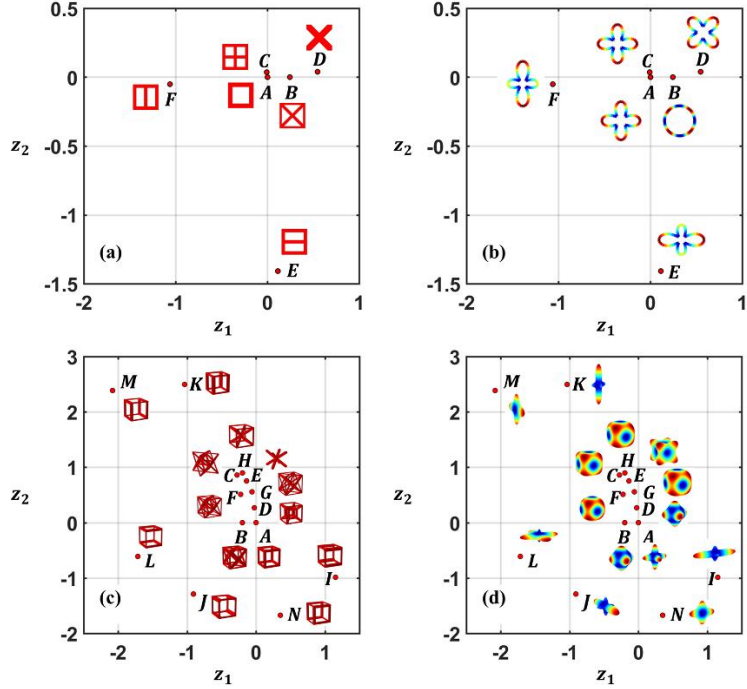

**FIGURE 6**: Latent spaces for 2D and 3D libraries, (a) and (b) are the latent spaces for the 2D library, while (c) and (d) are the latent spaces for the 3D library. Different classes of microstructures are marked with their geometries and elasticity modulus surfaces.

In the 2D case, classes A through D with cubic symmetry cluster on the upper right corner of the latent space. This cluster and the other two classes (E and F) are relatively distant from each other, which is consistent with their differences in the directional characteristics of the stiffness matrix. Within those cubic symmetric classes, A and C are close to each other in the latent space, though they have different topologies. This result makes sense because the latent space is constructed based on the similarities between different microstructures' property responses. In the latent space of the 3D library, different classes form a pair of concentric rings in the latent space, with cubically symmetric classes A through H on the inner ring and solely orthotropic classes I through N on the outer ring. This is because these two symmetry types have distinct stiffness matrices illustrated by their modulus surfaces.

From these examples, we conclude that MR-LVGP modeling provides substantial insights and easy interpretations of the characteristics of different microstructure classes, inducing an interpretable distance metric between different microstructure concepts.

Compared with other representations in machine-learning-based techniques, such as integer encoding and one-hot encoding, this organized latent space representation is more desirable because: a) complex correlation can be expressed with a much lower dimensionality and a more condense embedding, and b) the distance between different vectors of design variables can encode the similarity information for optimization. Another desirable feature of our MR-LVGP model is that the stiffness matrix can change in a smooth and continuous way by varying latent variables. Taking the latent space for the 3D library in Figure. 6 (d) as an example, when we examine the shape of the elasticity surface located on the inner ring in a clockwise direction starting from A, we observe that it begins with a star-like shape, gradually expands into a sphere, followed by a cube with "antennae", and then transforms back to the beginning in a reversed order. A similar cycle with this smooth transition can also be identified on the outer ring.

Moreover, we find that this kind of transition exists not only on certain tracks but also in the whole latent space. For demonstration, we fix the volume fraction to be 0.5 and then sample the latent space of the obtained MR-LVGP to get the spatial distribution of the multi-response elasticity modulus surfaces as shown in Figure. 7.

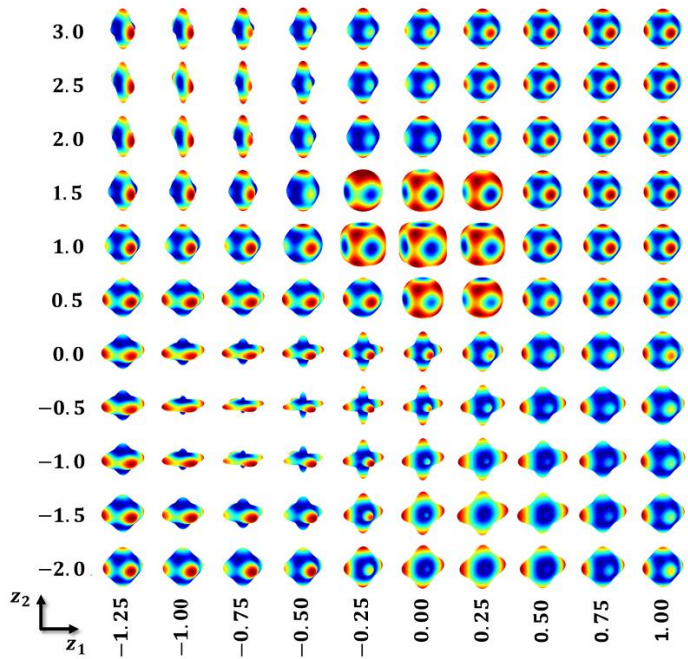

**FIGURE 7**: A set of elasticity modulus surfaces obtained by sampling on the latent space of 3D microstructures with the volume fraction fixed to 0.5

It is noted the modulus surfaces can change smoothly following any continuous route in the latent space, enabling the extraction of gradient information. This characteristic is critical for the integration of MR-LVGP models into multi-scale topology optimization explained in the later sections.

The concept of a low-dimensional latent space is also featured in some dimension reduction and deep generative methods, e.g. GPLVM [49, 50], GAN [51], VAE [52] and deep Gaussian processes [53]. However, fundamental differences exist between our method and these latent-variable models, with distinct functions and input/output definition. Specifically, both dimension reduction and deep generative methods generally create unsupervised learning models. They are geared to reconstruct or generate high-dimensional samples from a low-dimensional space with no direct link to the response prediction. Also, most deep generative models construct a latent space based on the features extracted from the high-dimensional geometric descriptor, e.g., pixelated matrix, rather than the qualitative variables used in this study. In contrast, our MR-LVGP model is a supervised predictive model with an aim to surrogate the relation between qualitative inputs and real-value response. It constructs a latent space of geometries based on the correlation in material responses, incorporating more physics into the learning model.

## 6. DATA-DRIVEN TOPOLOGY OPTIMIZATION

### 6.1. Integration of MR-LVGP into Multiscale TO

MR-LVGP models map different microstructure design concepts into a low-dimensional and continuous latent space and fully capture the collective effect of microstructure class and volume fraction on the stiffness matrix. With the meaningful distance metric, continuity, and gradient information, it is straightforward to replace the nested microstructure designs and homogenization process in multiscale TO with the fitted MR-LVGP models for higher efficiency.

Denote the fitted MR-LVGP model for the microstructure library as $Y(\rho, z(t))$, where $\rho$ is the volume fraction of microstructures (a smaller $\rho$ corresponds to microstructures with thinner rods as illustrated in Section 5.1), $t$ is the class of microstructures, $z = [z_1, z_2]^T$ is a 2D vector of latent variables corresponding to $t$, and $Y$ is a vector of independent components of the stiffness matrix. In multiscale TO, the full structure is divided into $N$ sub-regions. We then associate each subregion with three design variables, $\rho$, $z_1$ and $z_2$, and obtain the corresponding stiffness matrix in each iteration as $\boldsymbol{k}(\rho, z_1, z_2) = \boldsymbol{k}(Y(\rho, z_1, z_2))$. Although the stiffness matrix can have a smooth transition when exploring the latent space, only those latent variables transformed from existing microstructures can have practical meanings. Therefore, we drive the optimization solutions to the predefined classes in the libraries by using the penalized stiffness matrix $\widetilde{\boldsymbol{k}}$:

$$\widetilde{\boldsymbol{k}}(\rho, z_1, z_2) = f(z_1, z_2)\boldsymbol{k}(\rho, z_1, z_2), \tag{16}$$

$$f = \exp\left\{-1/\gamma \cdot \min_t(\|\boldsymbol{z} - \boldsymbol{z}(t)\|_2^2)\right\}$$
$$= \max_t\{\exp(-1/\gamma \cdot \|\boldsymbol{z} - \boldsymbol{z}(t)\|_2^2)\}, \tag{17}$$

where $f: R^2 \to (0,1]$ is the penalty function and $\gamma$ is a decay parameter of the penalty. To put it in an intuitive way, this penalization will make the mechanical properties decay with the nearest distance to the set of latent variables corresponding to existing classes in the library. To integrate it with TO, we adopt an approximated differentiable penalty function to replace the maximization operator:

$$f = 1/\lambda \cdot \ln\left\{\sum_t \exp\left(\lambda \cdot \exp(-1/\gamma \cdot \|\boldsymbol{z} - \boldsymbol{z}(t)\|_2^2)\right)\right\}, \tag{18}$$

where $\lambda$ is a large constant. Based on our experience, we recommend $\lambda$ to be 500 and $\gamma$ to be the diagonal length of the minimum bounding rectangle for the discrete latent variables. The multiscale TO problem can then be formulated as

$$\min_{\boldsymbol{\rho}, \boldsymbol{z_1}, \boldsymbol{z_2}} c(\boldsymbol{\rho}, \boldsymbol{z_1}, \boldsymbol{z_2}) = \mathbf{U}^T \mathbf{K} \mathbf{U} = \sum_{e=1}^N \boldsymbol{u}_e^T \widetilde{\boldsymbol{k}}_e\left(\rho^{(e)}, z_1^{(e)}, z_2^{(e)}\right)\boldsymbol{u}_e,$$
$$s.t. \ \mathbf{K}\mathbf{U} = \mathbf{F},$$
$$V(\boldsymbol{\rho}) \leq V_{max}, \tag{19}$$
$$z_i^- \leq z_i \leq z_i^+, i = 1,2,$$
$$0 < \rho_{min} \leq \rho \leq \rho_{max},$$

where $\mathbf{K}$ is the global stiffness matrix, $\mathbf{U}$ and $\mathbf{F}$ are global displacement and loading vectors respectively, $\boldsymbol{u}_e$ and $\widetilde{\boldsymbol{k}}_e$ are elemental displacement and stiffness matrix respectively, $V$ and $V_{max}$ are the solid material volume fraction and its upper constraint respectively, $z_i^-$ ($z_i^+$) is the lower (upper) bound for the $i^{th}$ latent variable, $\boldsymbol{\rho}_{min}$ is a vector of small values to avoid singularity and $\boldsymbol{\rho}_{max}$ is a vector of the maximum volume fraction for each subregion.

For this optimization problem, the sensitivities of the objective function $c$ with respect to the design variables of each microstructure can be obtained through the adjoint method and chain rule as

$$\frac{\partial c}{\partial \rho^{(e)}} = -f\left(z_1^{(e)}, z_2^{(e)}\right) \boldsymbol{u}_e^T \sum_i \frac{\partial \boldsymbol{k}_e}{\partial Y_i} \frac{\partial Y_i}{\partial \rho^{(e)}} \boldsymbol{u}_e, \tag{20}$$

$$\frac{\partial c}{\partial z_j^{(e)}} = -\boldsymbol{u}_e^T \left\{ f\left(z_1^{(e)}, z_2^{(e)}\right) \sum_i \frac{\partial \boldsymbol{k}_e}{\partial Y_i} \frac{\partial Y_i}{\partial z_j^{(e)}} + \frac{\partial f}{\partial z_j^{(e)}} \boldsymbol{k}_e \right\} \boldsymbol{u}_e, \tag{21}$$

where $\frac{\partial Y_i}{\partial \rho^{(e)}}$ and $\frac{\partial f}{\partial z_j^{(e)}}$ can be obtained through direct differentiation of (15) and (18), respectively.

With the above definition of the optimization problem and sensitivities information, we propose a sequential three-stage method to obtain the optimized multiscale structure with multiclass microstructures. The flowchart for this is shown in Figure. 8.

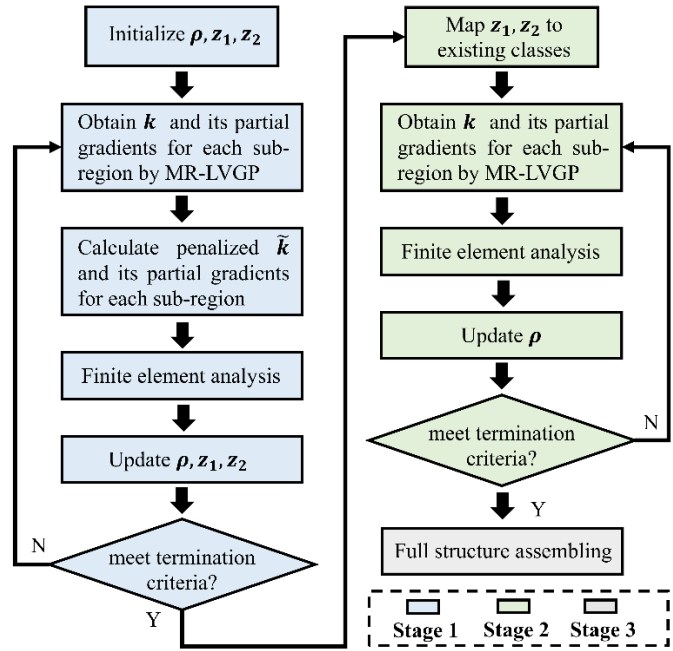

**FIGURE 8**: Flowchart for the data-driven multi-scale topology optimization

In Stage 1, both quantitative volume fraction and latent variables representing the class of microstructures are taken as design variables. The optimization problem (19) is solved by iteratively updating the design variables $\boldsymbol{\rho}, \boldsymbol{z_1}$ and $\boldsymbol{z_2}$ with the method of moving asymptotes (MMA) [54] based on the sensitivity calculated from (20) and (21). The optimization will terminate when the change in design variables (normalized) is less than 0.01 or the number of iterations exceeds 200. In this paper, the initial design is set to be a full structure consisting of the first class of microstructure with the same volume fraction $V_{max}/V_0$, where $V_0$ is the overall volume for the design space. Note that the penalization can drive the latent variables to those discrete points mapped from existing classes of microstructures but an exact convergence is not guaranteed. The possible resultant "intermediate classes" cannot be used to reversely generate corresponding topologies. In this case, the optimization result may not be accurate due to the penalization, which is similar to the issue of intermediate density values in the classical SIMP.

This issue is addressed in Stage 2. Specifically, the $\boldsymbol{z_1}$ and $\boldsymbol{z_2}$ results from Stage 1 optimization will be mapped to the nearest classes of microstructures in the latent space. The same optimization procedure in Stage 1 will be repeated in Stage 2 but with only volume fraction $\boldsymbol{\rho}$ as design variables. In Stage 3, the optimized structure of the microstructure for each subregion can be determined from the result of Stage 2 to assemble the full structure with the optimized performance.

### 6.2. Design Applications

In this section, we apply the proposed method to a few classical multiscale TO problems in both 2D and 3D cases. The design results will be compared with the optimized designs using a single type of microstructure to demonstrate the advantages of using a multiclass library. For all design cases, the matrix material is assumed to have relative Young's modulus 1.0 and Poisson's ratio 0.3. A filtering technique is applied to both quantitative and latent variables to avoid checkerboard patterns and excessive local flipping of the microstructure class.

To better illustrate the high adaptability to spatially varying stress conditions, we deliberately choose an L-shape optimization problem with distinctly different stress distributions in different parts of the structure as shown in Figure. 9.

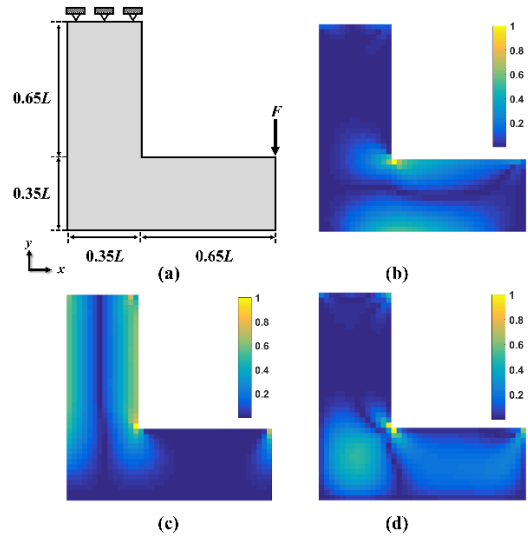

**FIGURE 9**: Example 1 – 2D single-loading L beam. (a) Problem setting for Example 1. The upper end of the L-shape structure is fixed. (b) The distribution of the normalized magnitude of horizontal stress. (c) The distribution of the normalized magnitude of vertical stress. (d) The distribution of the normalized magnitude of shear stress.

Specifically, the L-shape beam is fixed on the top while a point force is loaded on the up-right corner of the lower rectangle. In this case, the stress in the vertical direction dominates the upper rectangular area while the bending region mainly bears shear stress. The stress distribution for the lower rectangular part is more complicated, with the outer part dominated by the horizontal normal stress and the inner region being shear-dominant. It should be pointed out that our algorithm only focuses on compliance minimization and does not directly take the stress into consideration. The discussion on the stress distribution here is only to demonstrate that different regions of the L-shape beam are under different deformation conditions, which results in spatially varying requirements for the magnitude and directional characteristics of the stiffness tensor to achieve lower overall compliance.

To obtain the optimized full structures, the beam is discretized into square sub-regions with a length of $0.025\ L$. In practice, the level of discretization will depend on the minimum feature size in manufacturing. We set the maximum overall material volume fraction to be $0.6$ and the maximum volume fraction for each subdomain to be $0.95$.

The full structures with single-class and multiclass microstructures are shown in Figure. 10 (a) and (b), respectively. For all single-class designs in this study, we only optimize the volume fraction distribution while the microstructures are fixed to be class A during the whole optimization process. We only show one microstructure in each subregion for the convenience of illustration. However, each sub-region could actually be tiled by multiple repeated microstructures to meet the homogenization assumption.

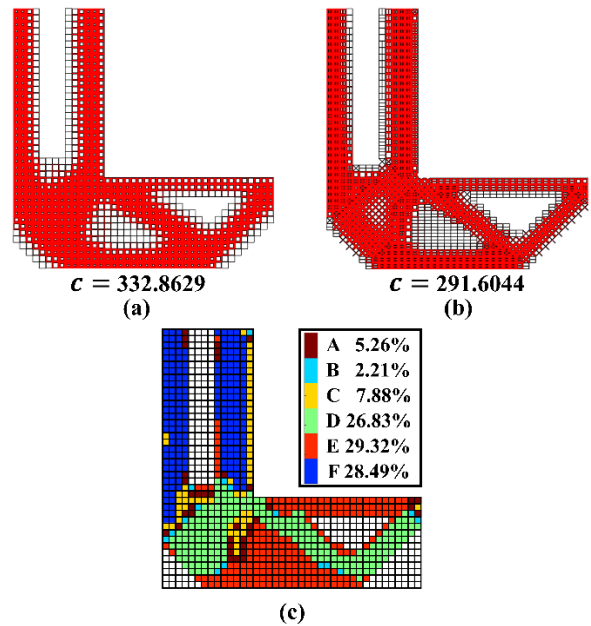

FIGURE 10: Design result of Example 1. (a) Full structure design with a single class of microstructures. (b) Full structure design with multiple classes of microstructures. (c) The distribution of different classes of microstructures in the multiclass design in (a), with the percentages of usage marked in the legend.

The achieved objective function $c$ (compliance) for the single-class design is 332.8629 while the value for the multiclass design is 291.6044. There are two reasons for this improvement in performance. The first reason is that different classes of microstructures can be allocated in a way that matches the principal stress direction. This can be indicated from the distribution of different classes of microstructures shown in Figure. 10 (c). In the multiclass design, microstructures in the upper rectangular area have thicker rods in the $y$ direction to bear the vertical loading while the bending region mainly contains microstructures with diagonal rods to resist the shear deformation. As expected, the outer layer of the lower rectangle prefers microstructures stiffer in the $x$ direction to resist the horizontal strain induced by bending while the inner region chooses microstructures with more diagonal rods. Another reason for the better performance is the better compatibility between the main load-bearing directions of macro- and micro-structures. Compared with the single-class structure, microstructures in the multiclass design have their main loading axes better conformed to the shape of the macro-structure.

The benefits of multiclass microstructures can also be indicated by the percentages of different classes used in the multiclass design. While no class of microstructure dominates the full structure, classes D through F rank top among all six classes in the library. This means that a better performance should be achieved by a combination of different microstructure design concepts. Therefore, using a single predefined microstructure design concept for the whole structure will be suboptimal for this and other general cases.

In terms of efficiency, compared with the physics-based multiscale TO, our data-driven approach replaces the numerous microscale design evaluations with a mixed-variable GP model obtained from the pre-computed library. Specifically, for this design example, assuming each microstructure has $100 \times 100$ elements and there are 924 subregions in the L-shape beam, there will be $924 \times 100 \times 100$ design variables in the original multiscale TO but only $924 \times 3$ (one quantitative variable and two latent variables for each microstructure) ones in our method. This treatment can avoid the time-consuming homogenization process and greatly reduce the number of topological design variables. Since there are only two extra latent variables associated with each subregion, the optimization process can have an efficiency comparable to the classical single-scale SIMP method. To further demonstrate the efficiency as well as the effectiveness of the proposed algorithm, we design a half Messerschmidt-Bolkow-Blohm (MBB) beam using our method for compliance minimization to compare with the optimal design obtained through $FE^2$ framework proposed by Xia et al. in [9], as shown in Figure 11.

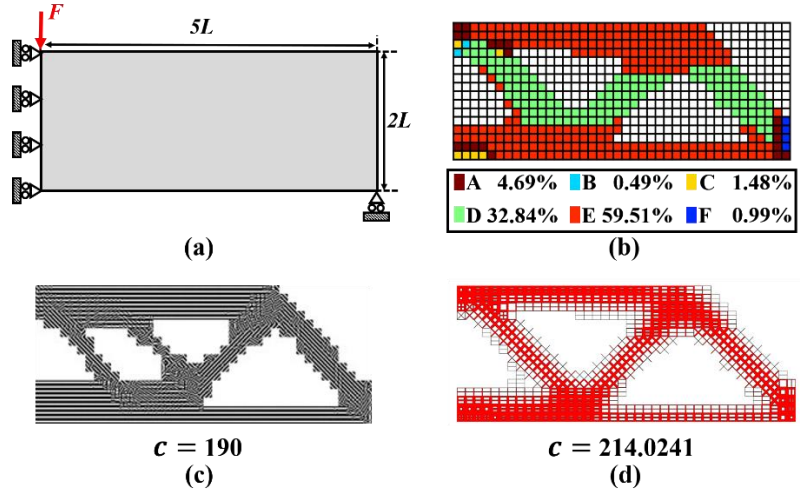

**FIGURE 11**: Example 2 – 2D single-loading half MBB beam. (a) Problem setting illustration. (b) The distribution of different classes of microstructures in the multiclass design shown in (d), with the percentage of usage marked in the legend. (c) Full structure design in Xia et al. [9] (d) Full structure design with multiple classes of microstructures.

We divide the design area into a $40 \times 16$ mesh and set the maximum overall material constraint to be 0.36, which is the overall material usage of the optimal structure obtained by Xia et al. in [9]. The optimized class distribution and the corresponding full structures are shown in Figure 11 (b) and (d). Compared with the optimal structure in Figure 11 (d) obtained by the $FE^2$ method, the compliance value of our optimal design is slightly higher, which is mainly due to the restriction of microstructure topology. However, the main load-bearing directions of the microstructures are similar in both designs. It takes the $FE^2$ method about 200 hours to obtain the assembled structure in this size while our data-driven method only takes fewer than five minutes.

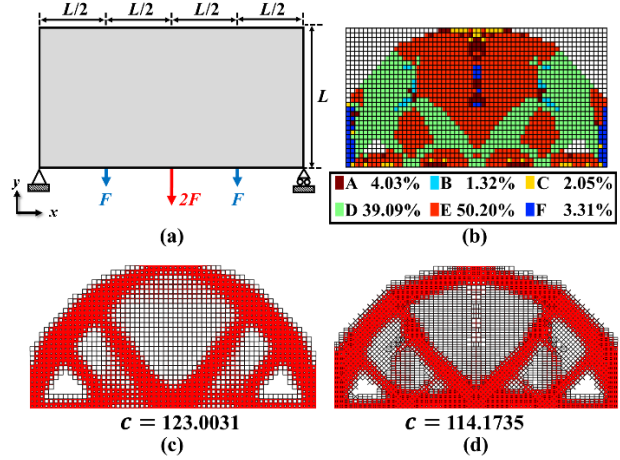

**FIGURE 12**: Example 3 – 2D multi-loading MBB beam. (a) Problem setting illustration. The beam is fixed on its low-left end and supported on the right. There are two sets of loading forces. The first one is loaded in the middle of the bottom layer and colored in red. The second one consists of two equal forces loaded at the two quarter points, which are colored in blue. (b) The distribution of different classes of microstructures in the multiclass design shown in (d), with the percentage of usage marked in the legend. (c) Full structure design with a single class of microstructures. (d) Full structure design with multiple classes of microstructures.

To demonstrate the benefit of multiclass microstructures in the multi-loading cases, we devise another numerical example depicted in Figure. 12 (a). The rectangular design space is divided into equal subregions with a resolution of $30 \times 60$. The objective function is the mean $c$ value for the MBB beam when applying the two sets of loadings respectively. We change the maximum overall material constraint to 0.5 and then obtain the optimized full structures as shown in Figure. 12 (b) through (d). The mean $c$ value is $123.0031$ for the single-class case and $114.1735$ for the multiclass case.

From the results, we observe the adaptive distribution of different classes of microstructures like before, aligning with the stress distribution. Compared with the first example, the percentage of class F in the full structure decreases significantly. This demonstrates that different design conditions have different required property distributions, which can benefit from the microstructure designs in the multiclass library. It is interesting to note that two designs have almost the same macro-structure, but the multiclass design can better match the main loading axes of microstructures with the shape of the macrostructure.

To demonstrate our method in 3D design, we optimize an L-shape structure similar to the 2D case with the maximum overall material distribution set to be 0.6. The beam is discretized into cubic sub-regions with a length of 0.1 L. The results are presented in Figure. 13 and Table 3, with the $c$ values being 485.6963 for single-class design and 351.7484 for multiclass design.

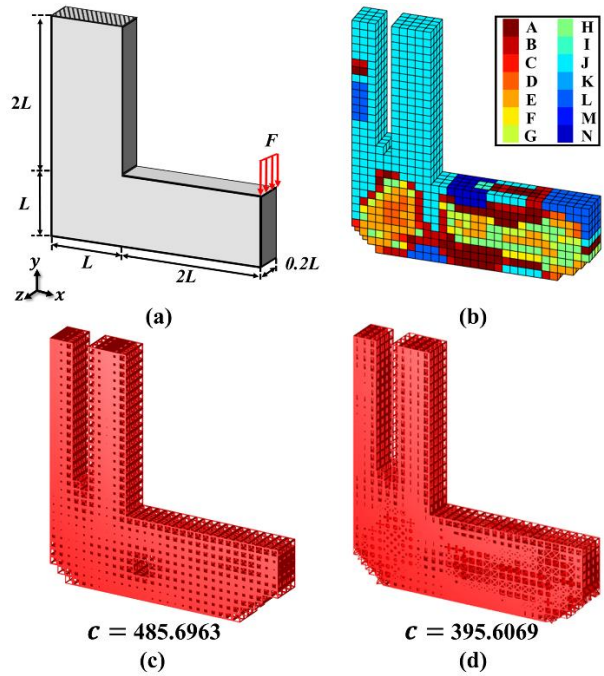

**FIGURE 13**: Example 4 – 3D single-loading L beam. (a) Problem setting. The upper end of the L-shape structure is fixed. (b) The distribution of different classes of microstructures in the multiclass design shown in (d). (c) Full structure design with a single class of microstructures. (d) Full structure design with multiple classes of microstructures.

**Table 3.** The percentages of different classes used in the multiclass full structure for Example 3.

| Class | Percentage | Class | Percentage |
|-------|------------|-------|------------|
| A | 12.17 | H | 8.22 |
| B | 9.56 | I | 0.44 |
| C | 1.22 | J | 38.17 |
| D | 5.33 | K | 0.00 |
| E | 11.50 | L | 6.00 |
| F | 2.50 | M | 0.00 |
| G | 3.33 | N | 1.56 |

It is noted that the main load-bearing direction of the microstructures in the 3D multiclass structure has a similar distribution as the one in the 2D case. The rectangular region connected to the fixed end prefers microstructure classes with thicker rods in the vertical direction. The outer layer of the rectangular area including the loading end has thicker rods in the horizontal direction. And the other regions need more shear-resistant microstructures with many diagonal rods.

From these design cases, it is evident that our MR-LVGP modeling and penalization techniques enable effective data-driven multiscale TO with multiple classes of microstructures. By using fewer design variables and avoiding the numerical homogenization process, we are able to optimize the full structure with much finer subregion division than physics-based methods with less computation cost. Compared with existing data-driven algorithms considering only single-class microstructures, each subregion in the full structure can selectively choose its class of microstructures to adapt to the local stress distribution. The method also provides better compatibility between the main load-bearing directions of macro- and micro-structures due to the use of predefined classes of microstructures, resulting in better design performance.

## 7. CONCLUSION

We propose a new multi-response LVGP model for mixed-variable datasets as well as a novel data-driven multiscale topology optimization (TO) method that can consider multiple classes of microstructures to design aperiodic multiscale structures efficiently. Compared with the conventional "free-form" multiscale TO methods, our proposed approach allows the consideration of a set of microstructure design concepts based on the existing knowledge of designers and consideration of manufacturability. The key idea is to map different types of microstructures into a continuous latent space using the proposed multi-response latent-variable Gaussian process (MR-LVGP) modeling method based on their effects on the multiple responses. By introducing a set of latent variables to represent qualitative inputs and a nonspatial covariance matrix of multiple responses, precise and computationally stable GP modeling is achieved for a mixed-variable dataset with high-dimensional responses. The original mixed-variable optimization problem for aperiodic multiscale structures can then be transformed into a continuous-variable one by including the latent variables as design variables and replacing the nested homogenization with our MR-LVGP model in the TO framework.

This MR-LVGP modeling approach has been applied to both 2D and 3D microstructure libraries to obtain a unified and continuous latent space for different classes of microstructures. A unique characteristic of MR-LVGP models is that the latent variables induce an interpretable distance metric reflecting the correlation between the mechanical responses of different classes. Microstructure design concepts with similar characteristics in properties, e.g. the directional characteristics of the stiffness matrix, will cluster in the latent space to form a well-organized pattern, enabling a clear visualization of the complex library. The interplay between different classes and properties of microstructures can be fully captured in the unified design space, which is a feature that suits the mixed-variable nature of material and structure designs. The fitted MR-LVGP models also enable the stiffness matrix to change in a smooth and continuous way when varying the latent variables. The multiscale TO with multiclass microstructures can then be realized with a simple modification of the classical SIMP method for TO which includes two-dimensional latent variables as extra design variables. Note that this proposed framework can be directly applied to other non-truss types of microstructures, though only truss-type designs are studied in this paper. This is because the latent space is directly related to the mechanical responses with no special requirement for the type of geometries. Provided that the microstructures are connected with each other to ensure manufacturability, domain knowledge can be utilized to freely choose types of microstructures included in the database.

The data-driven multiscale TO is applied to both 2D and 3D design cases. With the precomputed library and significantly reduced amount of topological design variables, the efficiency of the data-driven multiscale TO method is comparable to the standard single-scale SIMP method. In all design cases, full structures with multiclass microstructures have better performance than those with single-class microstructures. This demonstrates the advantages of aperiodic structures with multiple microstructure patterns, which can couple the macro- and micro-designs to better match local stress distributions in more general cases.

For future works, our method will be applied to multi-physics cases, such as the heat conduction structure design. Compared with the simple compliance minimization problem, multi-physics design may benefit more from the spatial varying property distribution. Multiscale optimization under loading uncertainty is another promising direction where a proper combination of different microstructure classes could provide more robust performance. Also, the homogenized properties might be imprecise in some cases because of the issues related to scale separation, which is a common challenge for multiscale TO. Currently, we use filtering techniques to avoid excessive local flipping of the microstructure types and assume each subregion is filled by numerous microstructures. In the future, we will explore the integration of our algorithm with reduced-order finite element methods to obtain more precise mechanical responses for microstructures. Some sophisticated techniques, such as the tuning of boundaries, will be included to ease possible stress concentration. Finally, even though the quantitative variables considered in this work are only associated with the volume fraction (density) and the design is focused on TO, the same proposed framework can be used to treat both materials properties and density as quantitative design variables for other material systems designs, realizing concurrent material and structure optimization.

**ACKNOWLEDGMENT**

The authors are grateful to Prof. K. Svanberg, from the Royal Institute of Technology, Sweden, for providing a copy of the MMA code for metamaterial designs. Support from the National Science Foundation (NSF) (Grant No. OAC 1835782) is greatly appreciated. Mr. Liwei Wang would like to acknowledge the support from the Zhiyuan Honors Program at Shanghai Jiao Tong University for his predoctoral study at Northwestern University.